\documentclass[fleqn,12pt,twoside]{article}
\usepackage{espcrc1}
\usepackage{epsfig}
\title{Experimental Conference Summary }
\author{Hans J. Specht\address{Physikalisches Institut, Universitaet Heidelberg,Philosophenweg 12, 69120 Heidelberg, Germany}}
\begin{document}
\maketitle

\section{ Introduction}
\label{intro}

Within the history of our field, this Quark Matter Conference has been truly unique, 
and I am delighted and grateful to the organizers for the invitation to summarize it. 
First, RHIC has arrived, and the very fact and the way it did will surely make this 
meeting unforgettable to all of us. Second, the comprehensive programs at the AGS and 
the SPS have (with one exception) come to an end, and it therefore seems appropriate 
now to also look back and critically assess the old and the new at the same time.

\begin{figure}[h]
\vspace{-.6cm}  
\begin{minipage}[t]{80mm}
\includegraphics*[width=9.4cm]{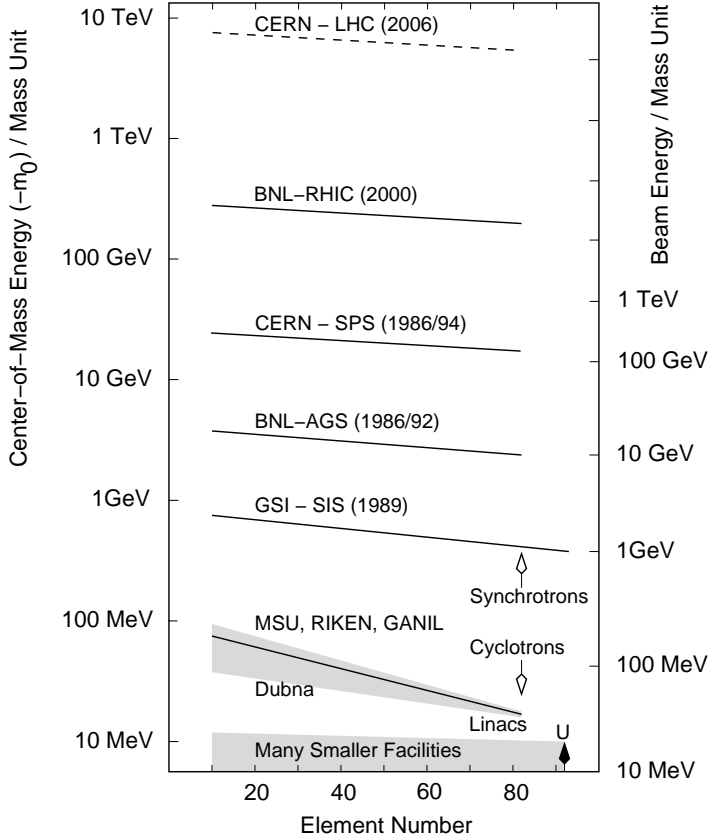}
\end{minipage}
\begin{flushright}
\begin{minipage}[t]{55mm}
\vspace{-10cm}
\caption{\footnotesize The major heavy ion accelerators worldwide. Nonrelativistic and relativistic, fixed 
target and collider facilities are put on a comparable basis by choosing the center-of-mass 
energy (minus rest mass) per nucleon for symmetric collision partners. Above 1 
AGeV, the spacing progressively increases even on a log-scale. The years indicated for 
each machine refer to first operation with light ions ($^{28}$Si, $^{16}$O/$^{32}$S) and 
very heavy ions (Au, Pb), resp. }
\label{fig:fig1}
\end{minipage}
\end{flushright}
\vspace{-2.4cm}  
\end{figure}

Indeed, I cannot resist to start even earlier. It is now about 20 years ago that the scene 
was set by a series of workshops in Berkeley 1979 \cite{1}, Darmstadt 1980 \cite{2} and, most 
important, Bielefeld 1982 \cite{3}, the start of the ``official" series of Quark Matter 
Conferences (later labelled ``2nd" to leave room for Darmstadt). In Bielefeld, particle 
physicists and nuclear physicists met for the first time in a systematic way 
to form a new community. Nearly everything was already there: first lattice 
results including the deconfinement transition, the chiral transition and indications for 
the soft equation of state, the astrophysical relevance, nuclear 
collision dynamics, most of the relevant observables including $\rho$-melting and hard 
probes like leptons and photons, the basic ingredients to the future experimental 
program (largely at the SPS) - and, most remarkably, the accelerator setting for the 
following 2-3 decades. As illustrated in Fig.~\ref{fig:fig1}, the preceding discussion on the LBL 
VENUS project [1]-[3] and the CERN ISR \cite{2,3} had been superseded by a quickly 
converging discussion on SIS, the AGS, the SPS and RHIC, the latter in Barton's 
paper \cite{3} still sailing under the traditional $pp$ project name ISABELLE (the AGS came 
in through the backdoor of that). Indeed, only the LHC was not yet born at that time. 

So now we celebrate the arrival of RHIC, delivering the highest center-of-mass energies ever made 
by mankind. The first day of the conference started with a grand firework.
There were charged multiplicities, high-$p_T$ spectra over many orders of magnitude, identified particles up 
to the exotica $\Omega^-/\Omega^+$, $\pi^\circ$-spectra, flow, HBT correlations, single 
electrons, and the 
announcement of tens of more detailed contributions in the parallel and poster sessions. 
At the end of the day I had a headache, and it was only in the course of the week and 
after the 4 reviews of the final morning that I came to fully appreciate what 
we were all witnessing during these days. Personally, I am not aware of any other 
example within our field or beyond, which produced such a rich spectrum of (albeit 
preliminary) new and exciting physics results in such a short time after the first run, at a 
brandnew machine with a set of brandnew elaborate and complicated experiments. I 
express my admiration, and I congratulate all of you, the management, the machine 
crew and the large number of enthusiastic young researchers who have made this 
miracle possible.

\section{ What have we learnt?}
\label{what}

\begin{figure}[htp]
\centerline{\includegraphics*[width=14.cm]{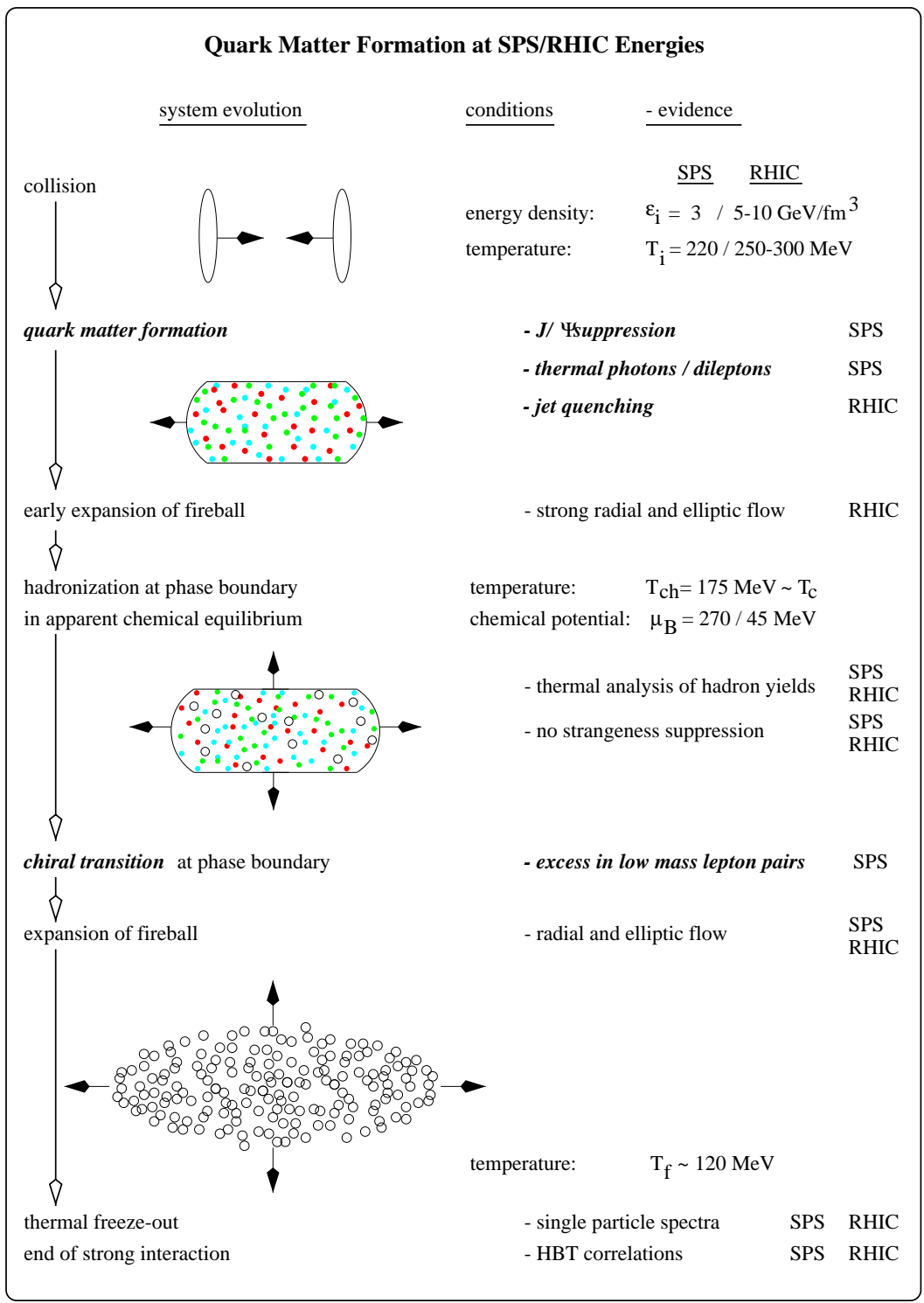}}
\caption{\footnotesize Fireball evolution following the impact of two heavy nuclei at ultrarelativistic 
energies. The various stages in the evolution are confronted with the experimental 
evidence and the experimentally determined conditions of the respective stage. Results 
(from mostly hard probes) related rather directly to quark matter formation are printed 
in {\it italics}. The labels SPS/RHIC refer to the energy regimes for which the 
respective evidence has thus for been observed.}
\label{fig:fig2}
\end{figure}

The substance of what I wish to discuss is shown in Fig.~\ref{fig:fig2} which can serve at the same 
time as an introduction, a guideline through the talk and a final summary. The figure is 
largely self-explanatory, confronting the fireball evolution after the collision impact with 
the experimental evidence for the 
interpretation and the conditions of the respective stage. With all due respect to RHIC, its first 
exciting results and its fascinating future potential, I will be courageous, use (what 
seems overdue) already at SPS energies the term ``Quark Matter" rather than the 
nebulous ``New State of Matter" of last year's CERN press release [4],
and systematically integrate the new information from RHIC 
as reported during this conference. 
We then arrive at the following sequence of events: 
Heavy Nuclei collide and reach initial conditions in terms of energy density and 
temperature well above the critical values for deconfinement. Quark Matter is formed as 
evidenced by the hard probes $J/\psi$ (SPS), high-$p_T$/intermediate mass photons/dileptons 
(SPS) and high-$p_T$ hadrons (RHIC). 
The fireball then expands under pressure (more at 
RHIC than at the SPS), hadronizes with parameters close to the expected phase 
boundary, possibly shows the influence of chiral restoration at that boundary as 
evidenced by low mass dileptons (SPS), strongly expands further under pressure and 
finally, after thermal freeze-out, ends as a cloud of non-interacting hadrons. From 
hadronization onwards, the fireball evolution appears to show essentially no difference 
between the SPS and RHIC.

\subsection{Initial Conditions}

\begin{figure}[h]
\vspace{-.5cm}
\begin{minipage}[t]{70mm}
\includegraphics*[width=10cm]{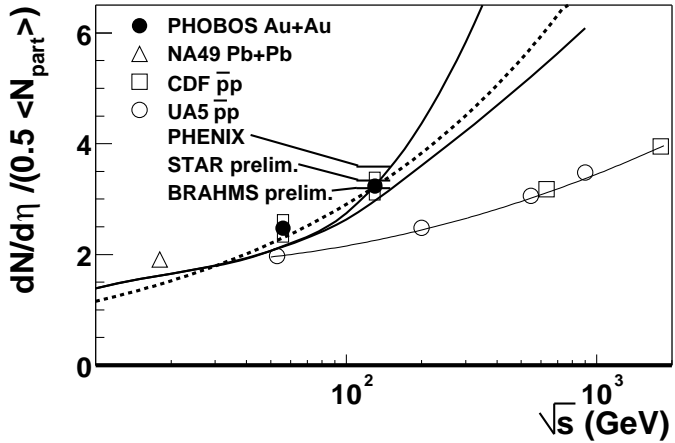}
\end{minipage}
\begin{flushright}
\begin{minipage}[t]{50mm}
\vspace{-7cm}
\caption{\footnotesize Dependence of $dN_{ch}/d\eta/0.5N_{part}$ for $|\eta|<1$ on the center-of-mass energy 
with results from all four RHIC experiments at $\sqrt{s_{NN}}$=130 GeV (from [6]).}
\label{fig:fig3}
\end{minipage}
\end{flushright}
\vspace{-2.6cm}  
\end{figure}

Charged particle multiplicities measured by PHOBOS were the first published results 
from RHIC \cite{5}. Fig.~\ref{fig:fig3}, taken from the review of Steinberg \cite{6}, shows 
the measurement of $dN_{ch}/d\eta/0.5 N_{part}$ as a function of $\sqrt{s}$ for all RHIC 
experiments in comparison to NA49 at the SPS and various $\overline{p}p$ data; for the 
interesting discussion on the onset of contributions from hard 
processes and the associated theoretical model lines in Fig.~\ref{fig:fig3} see ref.~\cite{6}. We 
use the fact that about 72 $\%$ more particles are produced at 
RHIC (at 130 GeV) than at the SPS and obtain, together with $<$$dE_T/dN_{ch}$$>$ = 0.8~GeV both at the 
SPS and at RHIC \cite{6}, initial energy densities of 3 and 5 GeV/fm$^3$, using the Bjorken 
formula \cite{7}. However, the hadron formation time $\tau$, taken as 1 
fm for this estimate, may well decrease as a function of $\sqrt{s}$ as proposed, e.g., by models of 
particle production based on parton saturation \cite{8}. If, conservatively, we allow for a 
factor of 2 at RHIC, we get a range 5-10 GeV/fm$^3$ and thus initial temperatures of 
T$_i$ = 220 and 250-300 MeV at the SPS and RHIC, resp., both well above the 
critical values of 173$\pm$8 (2 flavors) and 154$\pm$8 (3 flavors) quoted by Karsch for lattice 
QCD during this meeting \cite{9}. This argument alone, used of course by many people over 
many years, confirms that the SPS had a genuine chance, irrespective of the much more 
favorable conditions at RHIC.

\subsection{ Quark Matter Formation}

The evidence for quark matter formation at SPS and RHIC energies is essentially based on 
hard probes. This does not 
imply that soft probes like strangeness would not also be of relevance; consistency with quark 
matter formation is just not sufficient if alternate 
explanations persist. I will discuss the hard 
probes in the order indicated in Fig.~\ref{fig:fig2}. 
 
{\bf $J/\psi$  suppression}

\begin{figure}[h]
\vspace{-1.6cm}
\includegraphics*[width=5.2cm]{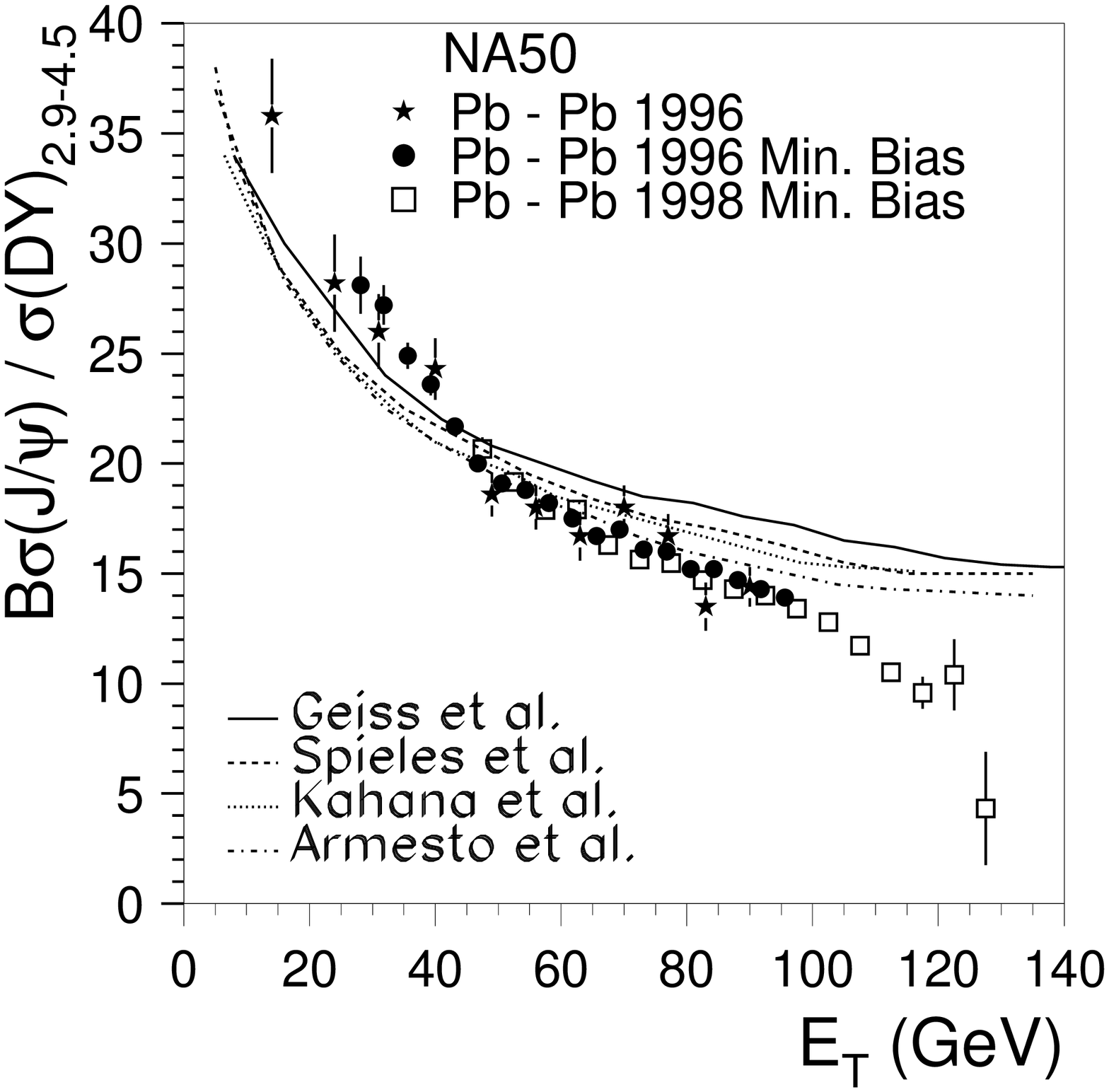}
\includegraphics*[width=5.2cm]{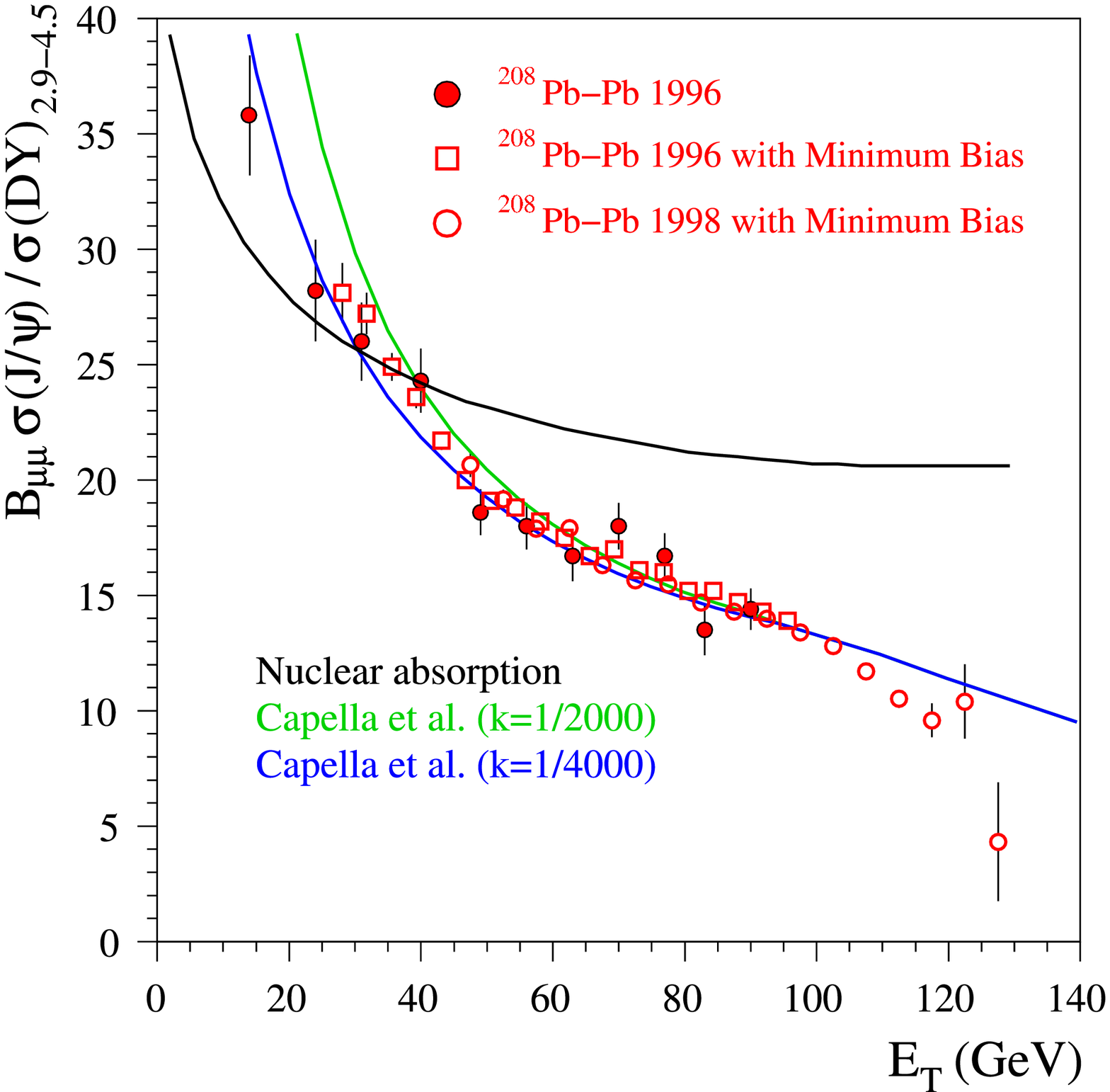}
\includegraphics*[width=5.2cm]{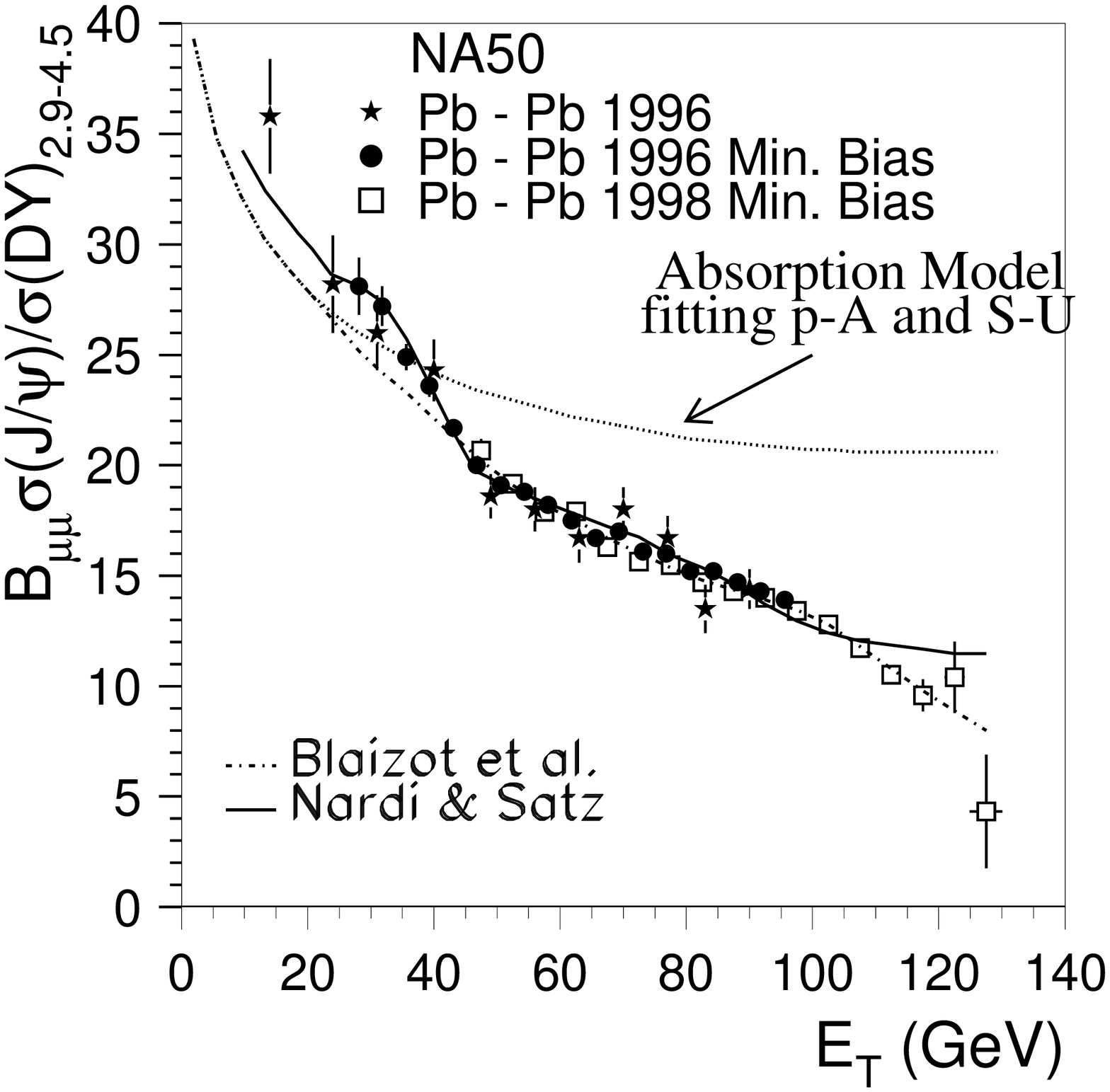}
\vspace{-.8cm}
\caption{\footnotesize Experimental results on the ratio $J/\psi$/DY compared to different models: suppression by 
hadronic comovers (left and middle) and by quark matter formation (right) (from [11] 
which also contains all model references).}
\label{fig:fig4}
\vspace{-0.9cm}
\end{figure}

The study of $J/\psi$ production, done by NA38/NA50, has been an integral part of the SPS 
program from the beginning. First signs for suppression were already reported after the 
very first data round (QM Nordkirchen 1987), when the appealing idea of melting in a 
deconfined phase as an existence proof for quark matter formation had just appeared 
\cite{10}. In the nearly 15 years since then, 100's of theoretical papers have challenged the 
unambiguousness of the idea with various hadronic suppression scenarios, driving a 
continuous feedback with better and better data systematics up to the present impressive 
level. The breakthrough came with PbPb collisions and the 
recognition of ``anomalous" suppression beyond the one understood as nuclear 
absorption. The controversy is still continuing, however, sometimes now not even free 
from irrationalism. Today's situation as summarized by Bordalo \cite{11} is illustrated in 
Fig.~\ref{fig:fig4}. The data are essentially those reported already at QM Torino 1999, while some of the 
model descriptions have only appeared more recently. Hadronic suppression based on 
conventional nuclear absorption is far off; hadronic suppression based on comovers and 
Qiu's \cite{12} new approach to nuclear absorption fail at least in shape. 
The assumption of 
deconfinement clearly gives the best description. The still controversial stepwise 
suppression \cite{13} with the apparent onset at $E_T\sim$ 40 GeV will be confirmed if true by the 
improved data at low $E_T$ taken in 2000, while the very natural idea 
of $E_T$-fluctuations at high $E_T$, coupled to complete suppression beyond a critical value of 
the local energy density \cite{14}, should perhaps be integrated with the onset- (or even 
stepwise-) suppression picture. New information on $J/\psi$ transverse momentum 
distributions was also presented \cite{11}; this still awaits critical modelling. Further 
experimental insight can be expected from NA60 which will e.g. look into the threshold 
behaviour (via a smaller collision system) and into the more open $\psi^\prime$ issue. In any 
case, $J/\psi$ suppression as it stands at this moment should be taken as the most compelling 
piece of evidence in favor of deconfinement which the field has produced thus far.

In retrospect, $J/\psi$ suppression at the SPS seems to be a fortunate accident of 
nature. Since only about 1 $\%$ of all charm produced appears in primary charmonia (the 
rest in open charm), it is conceivable that $J/\psi$'s can also newly be created at the 
hadronization stage (see 2.3). Surprisingly, investigations along this line have only 
recently been done \cite{15,16}, concluding that, although genuine thermal charm production 
is small, direct charm production by hard processes may lead to a level of statistically 
produced $J/\psi$'s masking the suppression mechanism for the 
primary $J/\psi$'s. At the SPS \cite{15}, 
the level does seem to come close, in particular if further charm enhancement is assumed 
(see below). However, the functional dependence on $E_T$ (or $N_{part}$) with its peculiar shape 
(name it different ``thresholds" or not) is not at all described, and the reduction relative 
to the primary level is in any case unchallenged. At RHIC \cite{15,16}, however, statistical 
production may be so effective as to lead to a net $J/\psi$ enhancement. That would indeed be 
spectacular, greatly help precision experiments and provide an alternative tool to study 
quark matter formation \cite{16} (if it does not suffer the fate of strangeness, hadronic phase space).

{\bf Thermal Photons/Dileptons}

\begin{figure}[h]
\vspace{-1.cm}
\includegraphics*[width=5.3cm]{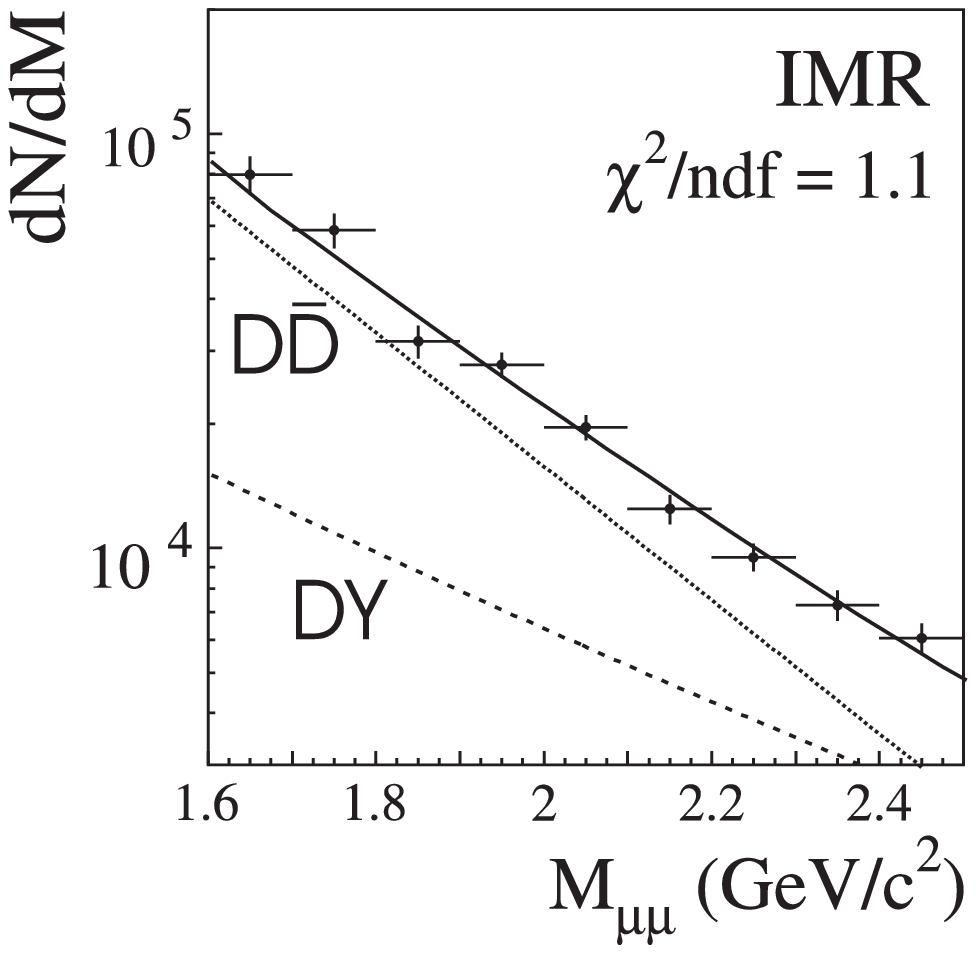}
\includegraphics*[width=4.9cm]{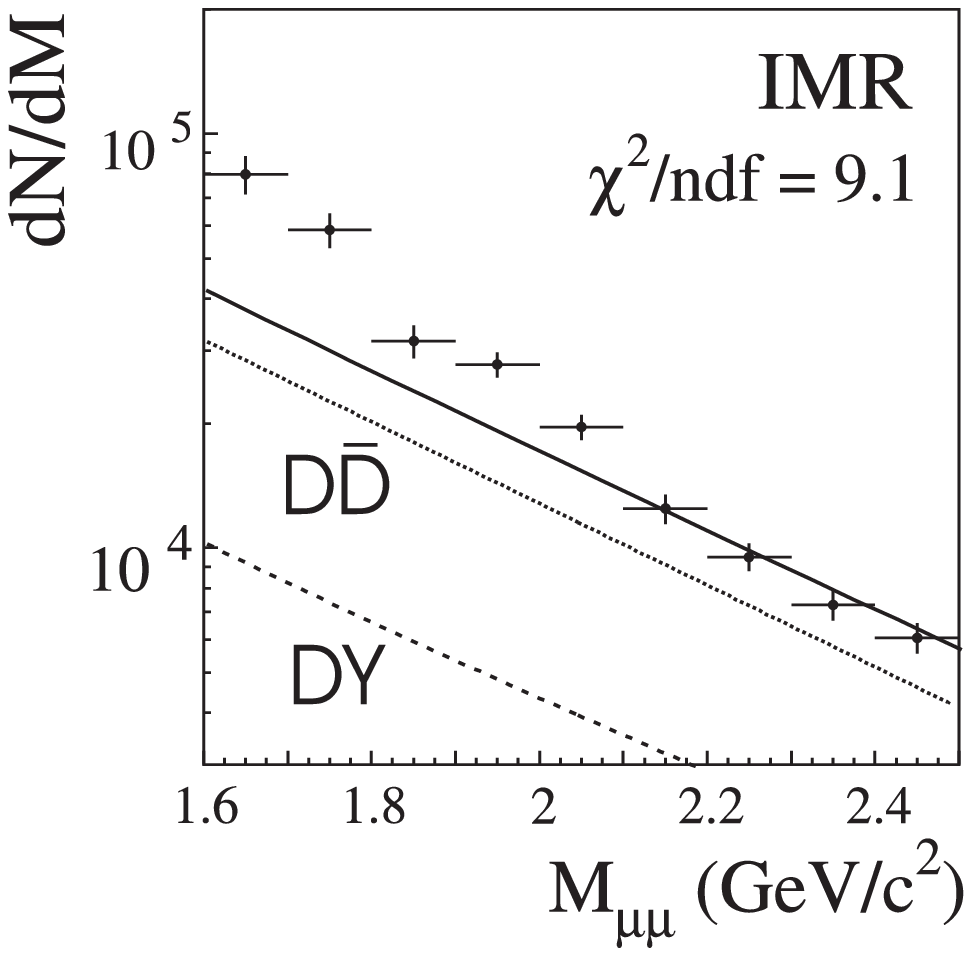}
\includegraphics*[width=5.45cm]{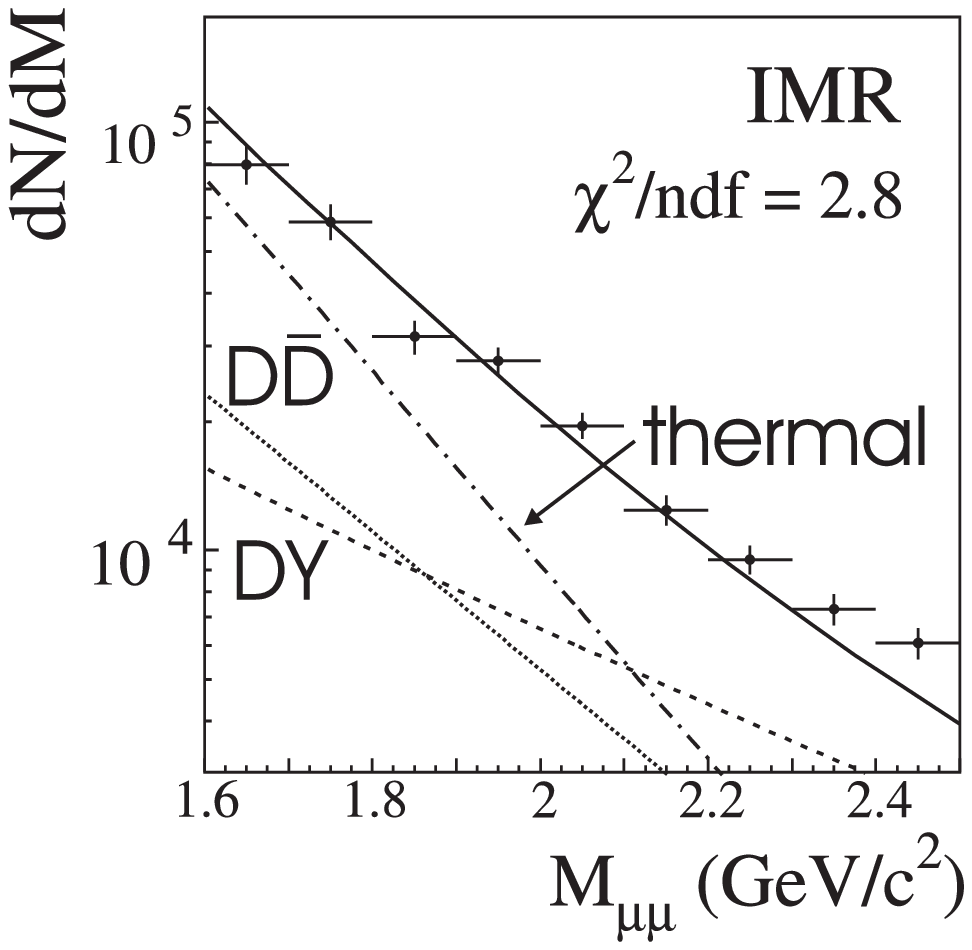}
\vspace{-1.1cm}
\caption{\footnotesize Dimuon mass spectrum in the intermediate mass region for central PbPb 
collisions compared to the prediction from different models: enhancement of charm 
production (left), D-meson rescattering (middle) and thermal radiation (right) (from [17]).}
\label{fig:fig5}
\vspace{-.4cm}  	
\end{figure}

Dileptons and direct photons are among the earliest observables proposed for quark 
matter diagnostics. Unfortunately, the experiments are extremely difficult, and even 15 
years after the start-up the situation has not even reached the level of relative maturity 
of the  $J/\psi$. Dimuons with intermediate mass, i.e. in the mass window between the $\phi$ and 
the $J/\psi$, have been measured by HELIOS 3 and, very systematically, by 
NA38/NA50, and results on an enhancement relative to the sum of Drell-Yan dimuons 
and simultaneous semileptonic decays of D and $\overline{D}$ mesons (or relative to 
properly scaled 
$pA$ which fits) have consistently been reported on several of the preceding Quark Matter 
Conferences. The present PbPb data from NA50 together with the options for 
interpretation are shown in Fig.~\ref{fig:fig5} \cite{17}. 
The data can be described either by artificially 
increasing the open charm yield, with a scaling factor rising with the number of 
participants up to a level of 3 for central collisions. If true, that would clearly be 
sensational, since thermal charm production has just been shown to be small \cite{15}, and 
production by hard processes is not easily modified towards such a dramatic 
enhancement. Alternatively (and more probably), the excess can be described by adding 
thermal radiation. The specific model used in Fig.~\ref{fig:fig5} \cite{18} is based on an expanding 
thermal fireball which explicitly includes an early (ideal) quark matter phase and a late 
hadronic resonance phase. The initial temperature required is about 190 MeV (not very 
sensitive up to 220 MeV), and the relative contribution from the deconfined phase is 
about 20-30 $\%$. Final confirmation of thermal radiation obviously requires an 
experimental determination of the level of open charm (of direct relevance also for 
statistical $J/\psi$ production, see above), and this is surely a primary motivation for the new 
NA60 experiment at the SPS.

\begin{figure}[h]
\vspace{-1.3cm}
\begin{minipage}[t]{90mm}
\includegraphics*[width=9.7cm]{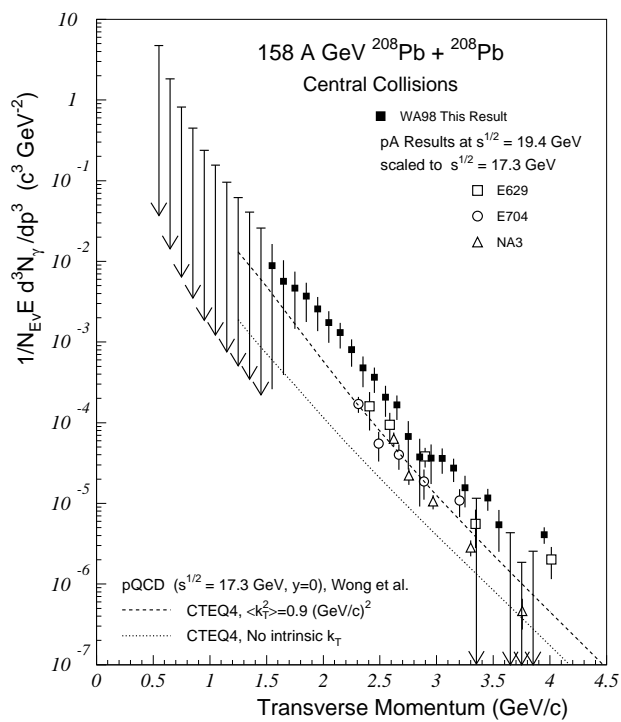}
\end{minipage}
\begin{flushright}
\begin{minipage}[t]{65mm}
\vspace{-7cm}
\caption{\footnotesize Invariant cross section for the production of real photons vs. transverse 
momentum. Data are from WA98 [19], the pQCD calculation from [21].}
\label{fig:fig6}
\end{minipage}
\end{flushright}
\vspace{-2.cm}  
\end{figure}

The measurement of direct photons is even more tough, due to the overwhelming 
background from $\pi^\circ$ and $\eta$ decay photons. All attempts with O and S beams, published 
by NA34/HELIOS, WA80 and NA45/CERES, have only resulted in upper bounds. The 
breakthrough came once again with PbPb collisions. The net photon $p_T$-spectrum, 
obtained by WA98 \cite{19}, is shown in Fig.~\ref{fig:fig6}. The model description of these data is 
presently quite open; an extensive review of the theoretical difficulties both in the high 
$p_T$ and the low $p_T$ region was given by Gale \cite{20} during this conference. The high $p_T$ 
part is usually believed to be dominated by hard QCD processes (like QCD Compton), 
somewhat analogous to Drell-Yan $q\overline{q}$ annihilation for high mass lepton pairs. 
The WA98 collaboration itself has added properly scaled $pA$ results in Fig.~\ref{fig:fig6} to 
argue that the PbPb 
data show an excess above hard processes up to very high $p_T$; indeed, the pQCD 
estimates \cite{21} contained in the plot support that. As a consequence, Srivastava's 
description of the data as thermal radiation \cite{22} requires initial temperatures of about 
330 MeV, unrealistically high compared to any other values obtained for SPS energies. 
However, various effects like nuclear $k_T$ broadening could add to increase the yield of 
hard processes at high $p_T$ \cite{20} suggesting, in lack of quantitative calculations, to fit the 
high $p_T$ part to hard processes with an ad hoc scaling factor. 
This is the approach taken 
by K\"ampfer \cite{22}, which reduces the $p_T$-region requiring an excess description by 
thermal radiation to $<$ 2.2 GeV. It is remarkable that the fireball model of this group is 
then able to describe the real photons of Fig.~\ref{fig:fig6}, the intermediate mass 
dimuons of Fig.~\ref{fig:fig5} 
and the low mass dielectrons of CERES (see below) with an identical set of parameters; 
the initial temperature required is 210 MeV, reasonably consistent with Rapp/Shuryak 
\cite{18} and the discussion of the initial conditions of section 2.1. All in all, direct photons 
obviously leave much room for improvements, and one can only hope that the situation 
at RHIC with higher initial temperatures will ultimately create a much more convincing 
case.

{\bf Jet Quenching}

\begin{figure}[h]
 \vspace{-3.cm}  
\begin{minipage}[t]{75mm}
\includegraphics*[width=8cm]{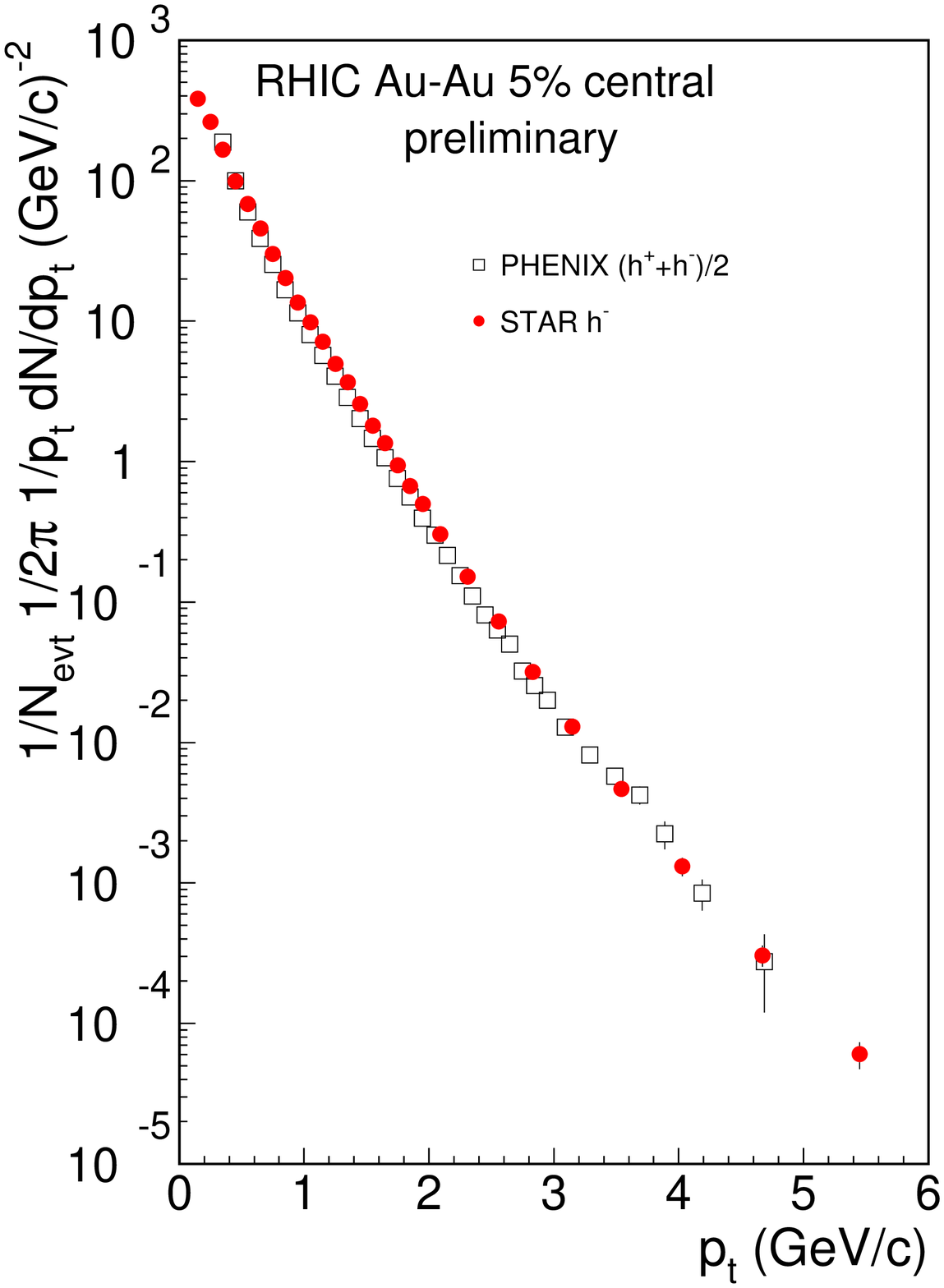}
 \vspace{-2.5cm}  
\caption{\footnotesize Inclusive cross sections for the production of negatively charged hadrons from 
STAR [25] and all charged hadrons from PHENIX [26] for the most central 5~$\%$ of the 
collisions. The data are independently normalized (from [29]).}
\label{fig:fig7}
%\end{figure}
\end{minipage}
 \begin{flushright}
\begin{minipage}[t]{80mm}
 \vspace{-12.5cm}
%\begin{figure}[h]
\hspace*{-1cm}\includegraphics*[width=10cm]{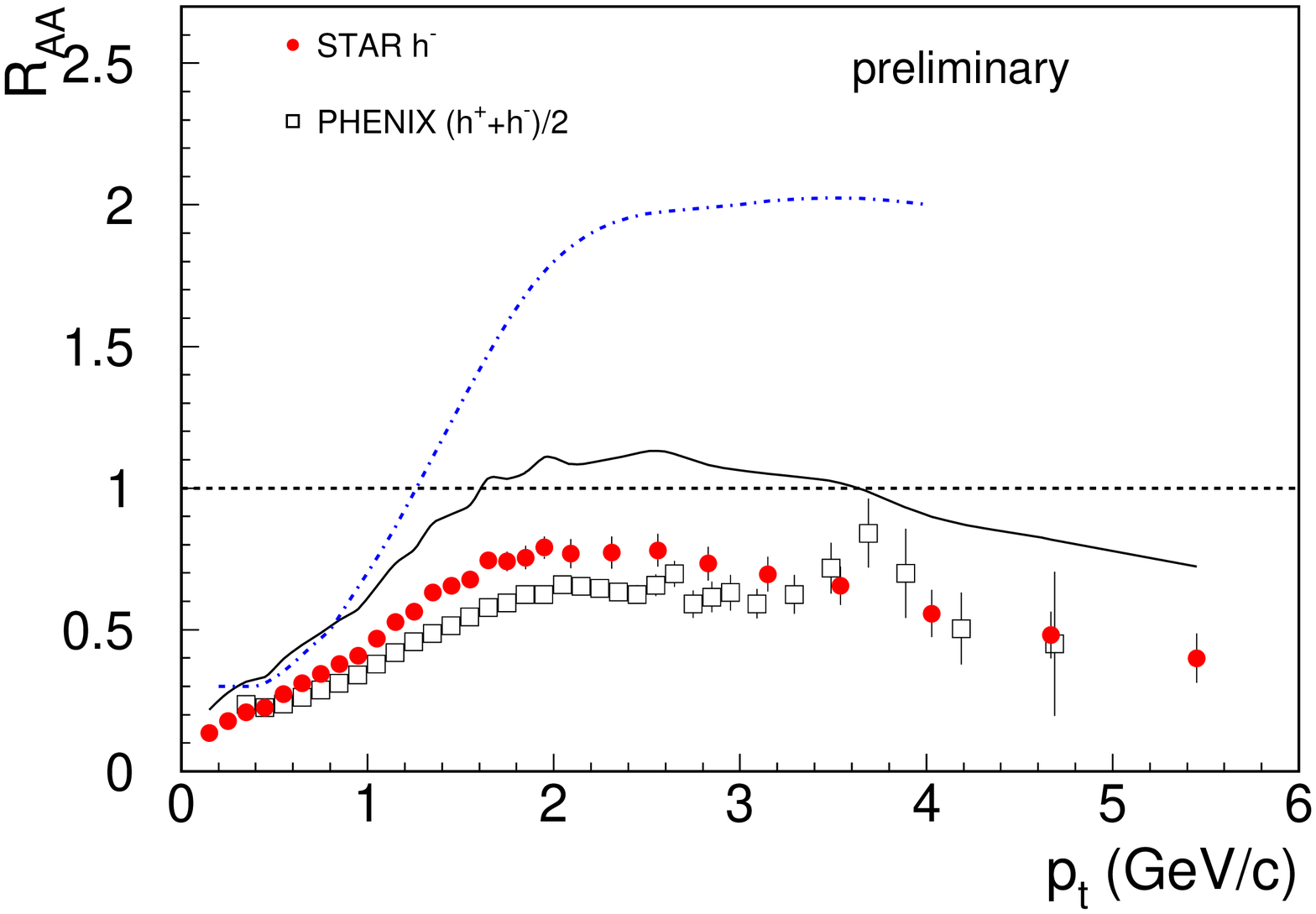}
 \vspace{-1.5cm}  
\caption{\footnotesize Data from Fig.~\ref{fig:fig7} normalized to nucleon-nucleon data 
with $<$$N_{binary}$$>$ = 945 (see text). The 
thin line is the upper limit of the systematical uncertainty. The dash-dotted line 
corresponds to the average of the SPS data from WA98, NA49 and NA45/CERES (from [29]).}
\label{fig:fig8}
\end{minipage}
\end{flushright}
\vspace{-1.3cm}  
\end{figure}

Out of all the impressive amount of new physics results from RHIC - the first evidence 
for jet quenching is {\it the} highlight of Quark Matter 2001! The idea, proposed about 10 
years ago \cite{24}, is quite simple. In the initial stage of the collision, quarks or gluons can 
scatter with high momentum transfer. The scattered partons, though fast, sense the hot 
and dense phase in the time-evolution of the fireball, loosing before escape a significant 
fraction of their momentum by induced gluon bremsstrahlung. 
The final fragmentation 
of the partons into jets of hadrons is then modified relative to the situation in free 
space, exhibiting reduced jet energies, i.e. reduced transverse momenta of the associated 
hadrons. Both STAR \cite{25} and PHENIX \cite{26}-\cite{28} have reported first results on 
significantly reduced inclusive hadron cross sections at high $p_T$, and a critical summary 
of the results including a comparison to SPS energies has been presented by Drees \cite{29}. 
Figs.~\ref{fig:fig7} and \ref{fig:fig8} repeat his essence. 
Fig.~\ref{fig:fig7} shows inclusive $p_T$ distributions for negatively 
charged hadrons from STAR and all charged hadrons from PHENIX. The agreement of 
the data over a range of 7 orders of magnitude is most impressive, illustrating the high 
level of analysis quality which these ``preliminary" data have already achieved.

To demonstrate jet quenching, one needs a comparison basis. This is provided by high 
energy $pp$ and $p\overline{p}$ data from CDF and UA1 (see \cite{29}), assuming that all 
high $p_T$ particle production in $AA$ (as in $pp$/$\overline{p}p$) results from binary hard 
collisions. A {\it nuclear modification factor} $R_{AA}$ can then be defined as 
$R_{AA}(p_T)=d\sigma_{AA}/dydp_{T}^{2}$/$<$$N_{binary}$$>$$d\sigma_{pp}/dydp_T^2$  \cite{29,30}, 
where the average number of 
binary collisions $<$$N_{binary}$$>$ is obtained from the inelastic cross sections and the nuclear overlap 
integral. Results on $R_{AA}$ as a function of $p_T$ for the two data sets of Fig.~\ref{fig:fig7} 
are shown in Fig.~\ref{fig:fig8}. Values $<$ 1 are expected for low $p_T$, since the cross 
sections in this region should 
scale with the number of participating nucleons rather than with the number of binary 
collisions. However, the high $p_T$ expectation of 1 for simple binary collision scaling is 
never reached, not to speak about the SPS level of 2 (due to the Cronin effect). Instead, a 
plateau is found at 0.6-0.8, followed by a decrease at still higher $p_T$. This is the evidence 
for jet quenching at RHIC. It finds support by the normalization of the central collision 
data to peripheral collisions rather than to $pp$ \cite{29}. It also finds support by the 
independent $\pi^\circ$ data of PHENIX [27]. These show a plateau value of only 0.4 suggesting 
that the reduction of $R_{AA}$ for the mixture of charged particles in Fig.~\ref{fig:fig8} is 
less radical than 
it would be for charged pions alone. Indeed, identified particle spectra from PHENIX [28] 
show the ratios $p/\pi^+$  and $\overline{p}/\pi^-$ to be unusually large, i.e. 
nearly 1 for $p_T >$ 2 GeV (possibly 
connected to the large radial  flow observed at RHIC, see 2.5 below), while a value of only 
0.2 is observed in $pp$. Accounting for the difference in (all charged)/$\pi$ between $AA$ and 
$pp$ would give a downward correction of a factor of $\sim$ 1.5 in Fig.~\ref{fig:fig8} and thereby 
consistency with the $\pi^\circ$ results.

Theoretically, the size of the observed effect can be accounted for by requiring an average 
energy loss of 0.25 GeV/fm for the scattered partons \cite{31}. It should be clear that this is 
only a phenomenological value averaged over the evolution history of the fireball. It does 
not separately reveal the specific energy loss of the partons and the characteristics and 
density of the medium the partons penetrated through.

\begin{figure}[h]
\vspace{-2.3cm}
\begin{minipage}[t]{80mm}
\includegraphics*[width=8cm]{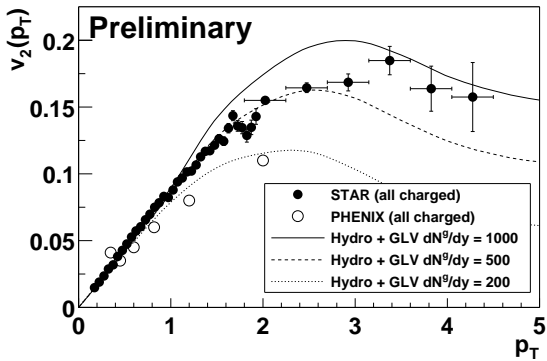}
\end{minipage}
 \begin{flushright}
\begin{minipage}[t]{60mm}
 \vspace{-6.5cm}
\caption{\footnotesize Azimuthal anisotropy v$_2$ relative to the reaction plane from STAR [32] and 
PHENIX [33]. The model calculations [35] combine hydrodynamics with hard scattered 
partons including an energy loss for 3 different initial gluon densities (from [6]). }
\label{fig:fig9}
\end{minipage}
\end{flushright}
 \vspace{-2.25cm}  
\end{figure}

In non-central collisions, the total parton propagation length should depend on the 
azimuthal direction. It is therefore conceivable that jet quenching would also show up as 
a specific azimuthal anisotropy of hadron spectra at large $p_T$, deviating from the low $p_T$ 
pattern. The suitable quantity to measure azimuthal anisotropy is v$_2$, the second 
Fourier coefficient of the azimuthal particle distribution relative to the reaction plane, 
usually called elliptic flow. Data on v$_2$ from STAR \cite{32} and PHENIX \cite{33} are shown in 
Fig.~\ref{fig:fig9}. The initial rise is consistent with hydrodynamics \cite{34,35}. The 
flattening observed 
by STAR for $p_T >$ 2 GeV/c, a clear deviation from hydrodynamics, is due to the onset of 
hard processes. The calculations contained in Fig.~\ref{fig:fig9} combine a soft hydrodynamic 
component with a hard pQCD component including jet quenching, i.e., a parton energy 
loss for different initial gluon densities \cite{35}. The middle curve describes the data 
reasonably well, the sensitivity of v$_2$ to the value of the energy loss is remarkable. 
Ultimately, the two manifestations of jet quenching , a depletion of high $p_T$ particle 
production and the flattening and decline of v$_2$, will require a consistent theoretical 
treatment with the same set of parameters.

All in all, jet quenching has given another exciting hint for quark matter formation 
already after the first round of experiments. Of course, there are and will be doubts as to 
the normalization procedure, the influence of radial flow, even hadronic scenarios. But 
there are enormous experimental reserves like a spectral extension up to 10 GeV and 
direct jet identification. One can only hope that the $J/\psi$ frustration of more than a decade 
does not repeat itself, and that ultimately numerical and quantitative information on the 
deconfined stage can be obtained.

\subsection{Hadronization}

Experimentally, the measurement of hadron yields and low $p_T$ hadron spectra belong to 
the easier part of the field, and a large amount of data was accumulated over the years. 
The model description of these data in terms of a statistical language was developed over the 
last decade by a number of authors, dating back, of course, to Hagedorn \cite{36} almost 
35 years ago. Reviews on hadron freeze-out have been given by 
Rischke \cite{37} and, with special emphasis on strangeness, by Redlich \cite{38} during this 
conference. I will follow present wisdom and distinguish {\it chemical} freeze-out, occurring 
earlier (at {\it higher} temperature) and determining particle abundances, from 
thermal-{\it kinetic freeze-out}, occurring later (at {\it lower} temperature) and determining particle 
momentum distributions. The former will be discussed in this section, the latter in 
section 2.5.

{\bf Global Particle Production}

The great news of this conference were the first particle yields from RHIC with 
contributions from all 4 experiments STAR [39-41], PHENIX \cite{42}, PHOBOS \cite{43} and 
BRAHMS \cite{44}; a summary of these data was given by Nu Xu \cite{45}. Compared to AGS and 
SPS energies, the most dramatic change concerns the central antibaryon/baryon ratios 
like $\overline{p}/p$, $\overline{\Lambda}/\Lambda$ and $\overline{\Xi}/\Xi$. 
The  $\overline{p}/p$ ratio rises from $<$0.1 at the SPS to about 0.6 (determined by 
all 4 experiments), implying that the system created at $\sqrt{s_{NN}}$ = 130 GeV is close 
to net-baryon free; the net-proton rapidity density for central collisions measured by STAR and 
BRAHMS is only about 10 \cite{41}. All this was of course anticipated, due to the loss of 
complete stopping at the much higher energies of RHIC. Conversely, the meson ratios 
like $K^-/\pi^-$ or  $K^-/K^+$ are found to be much closer between 
the SPS and RHIC.

\begin{figure}[h]
\vspace{-.8cm}
\begin{minipage}[t]{70mm}
\hspace{-.7cm}\includegraphics*[width=11cm]{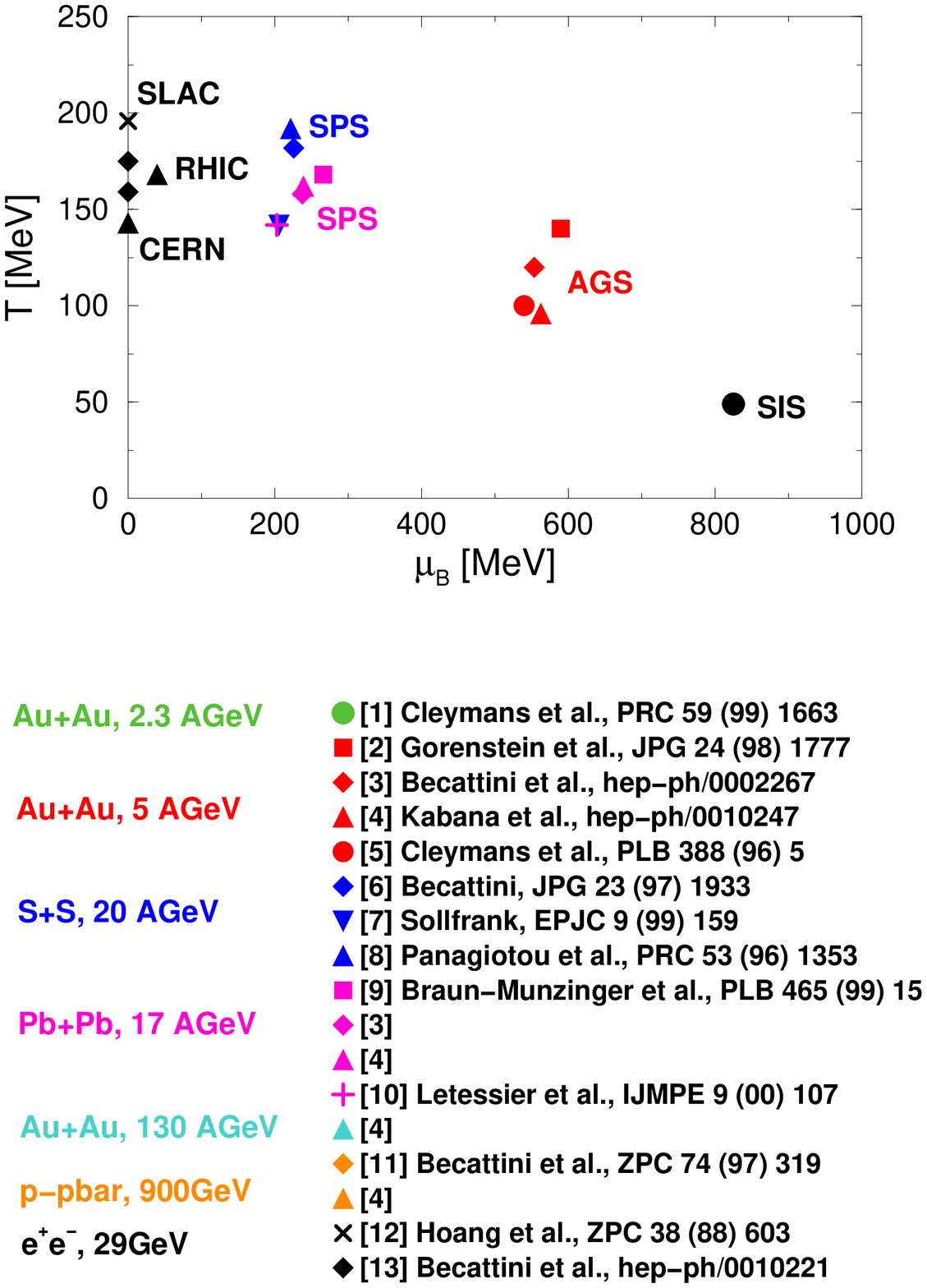}
\end{minipage}
\begin{flushright}
\begin{minipage}[t]{65mm}
\vspace{-7.8cm}
\caption{\footnotesize Systematics of the parameters T and $\mu_B$ extracted from measured hadron ratios 
for $e^+e^-$, $p\overline{p}$ ( at $\mu_B$=0) and nuclear collisions at RHIC, SPS, AGS and SIS. The average 
of the data points corresponds to a systematics for which energy density / total particle 
density $\sim$ 1 GeV [48]. The 12 references for the individual points are contained in [37] in these 
proceedings.}
\label{fig:fig10}
\end{minipage}
\end{flushright}
\vspace{-2.5cm}
\end{figure}

The statistical model commonly used to describe particle production at chemical freeze-out 
for $AA$ collisions is formulated in the {\it grand canonical} ensemble with global baryon, 
strangeness and charge conservation. All particle ratios are then a function of only two 
independent parameters, the temperature T and the baryon chemical potential $\mu_B$. 
A compilation of the parameters T and $\mu_B$  required to 
describe measured particle production for $AA$ collisions at RHIC, SPS, AGS and SIS 
energies is shown in Fig.~\ref{fig:fig10} \cite{37}; additional RHIC values can be found in \cite{45,46}. The 
reasons to choose this particular compilation rather than others [38, 47-49] are several 
fold: a demonstration of the long list of authors presently contributing, an illustration of 
the systematical errors of the parameters as visible by the scatter of the individual 
points, and a (somewhat incomplete) inclusion of particle production in elementary 
reactions like $p\overline{p}$ and $e^+e^-$. Two features are noteworthy. First, the averages 
of the points 
can be connected by a common line of constant energy per 
particle, $<$$E$$>$$/$$<$$N$$>$$\sim$ 1 GeV \cite{48}. This is 
of great interest in itself, implying the existence of an energy scale below which inelastic 
collisions stop (see \cite{48} for interpretation), but there is no connection to quark 
matter formation which is the main issue of this summary. Second, and that {\it is of} 
relevance for the issue, the temperature values for the SPS, RHIC and the elementary 
reactions are essentially identical and numerically, within errors, equal to the critical 
value $T_c$ for deconfinement from lattice QCD. How close the numbers are can best be 
illustrated (with smallest systematical errors) by quoting the values (in MeV) from only 
two groups of authors (which even work together): 168$\pm$10 \cite{50}, 175$\pm$7 \cite{51}, 
166$\pm$6 \cite{52} for the SPS, 175$\pm$7 \cite{46} for RHIC, 169$\pm$2 for $pp$ at 
$\sqrt{s}$ = 27 GeV [53], 175$\pm$15/170$\pm$12 for $p\overline{p}$ 
at $\sqrt{s}$ = 200/900 GeV [53] and 169$\pm$4 (revised)/167$\pm$2 for $e^+e^-$ 
jet fragmentation at $\sqrt{s}$ = 29/91 GeV (PEP-PETRA/LEP) [54].

This surely cannot be fortuitous. To the extent that in $e^+e^-$ hadrons are born from a 
preceding $q\overline{q}$ pair - could there be any better evidence, that in $AA$ at SPS or RHIC 
energies hadrons are also born from an ensemble of preceding partons? Is it the basic 
characteristics of string fragmentation which sets the scale for the universal 
hadronization parameter ``T" $\sim$ 170 MeV? So universal indeed that it also describes the 
momentum scale of soft particle production in $pp$ or in $e^+e^-$ (orthogonal to the jet axis)? Is 
the notion of chemical equilibrium among hadrons really appropriate? It is odd for jet 
fragmentation, and it suffers from the internal inconsistency in $AA$ that the medium 
effects on hadron masses and decay widths which are {\it known} to exist at the stage of 
hadronization, are not incorporated (if they were, T would very much 
decrease \cite{55}). Is it not more appropriate, as Rischke \cite{37} reminded us, to look at 
multiparticle production as saturating the available phase space (``born into apparent 
equilibrium")? And finally - is it some basic feature of QCD which we need to understand, that 
the scale parameter T in hadronization and the critical temperature T in 
thermodynamics are numerically close or even identical? And both close to the basic 
scale $\Lambda_{QCD}$?

{\bf Strangeness Production}

\begin{figure}[h]
\vspace{-.6cm}
\begin{minipage}[t]{85mm}
\includegraphics*[width=7.4cm]{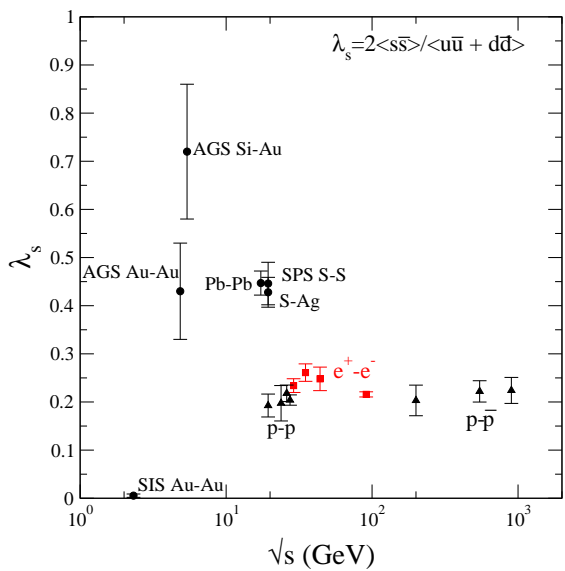}
\end{minipage}
 \begin{flushright}
\vspace{-7cm}
\begin{minipage}[t]{65mm}
\caption{\footnotesize Strangeness suppression factor $\lambda_S$ as a function of nucleon-nucleon center-of-mass energy $\sqrt{s}$. Values for $AA$ collisions are from [52], values for elementary reactions have been taken from [59] (from [52]).}
\label{fig:fig11}
\end{minipage}
 \end{flushright}
\vspace{1.5cm}
\end{figure}

Strangeness enhancement relative to elementary reactions like 
$pp$ or $e^+e^-$ has been proposed as a signature for quark matter formation almost 20 years 
ago \cite{56}. Experimental evidence for enhancement has also been with us since long, 
culminating in the huge factor of 17 for the triple strange hyperon $\Omega$ as measured by 
WA97 and reported again from NA57 during this conference [57]. A compact and elegant way 
to illustrate the relative level of strangeness production in $AA$ collisions and elementary 
collisions is provided by the strangeness suppression 
factor $\lambda_s = 2$$<$$s\overline{s}$$>$$/ ($$<$$u\overline{u}$$>$$+$$<$$d\overline{d}$$>$$)$ 
\cite{58}, measuring the multiplicity ratio of newly created valence quark-antiquark pairs (before 
resonance decays). A recent compilation of the $\lambda_s$-systematics is shown in 
Fig.~\ref{fig:fig11} \cite{52}. The 
elementary reactions reach a level of 0.2-0.25, rather independent of $\sqrt{s}$. Nuclear 
collisions, on the other hand, lie higher by a factor of 2, similar for the SPS and the 
uppermost energy of the AGS. Unfortunately, RHIC points have not yet become 
available, but it would be a surprise if they would show a huge difference to the SPS. 

Despite enormous efforts, the interpretation of the difference in strangeness production 
between nuclear and elementary reactions has continued to be controversial and 
inconclusive, up to the time of this conference as illustrated by Redlich \cite{38}. It is clear 
since some years that the statistical hadronization approach describes hadrons with 
strangeness just as near-perfect as all other hadrons, implying that the term 
``strangeness suppression" (referred to elementary reactions) as commonly used in 
particle physics may be more appropriate than ``strangeness enhancement". 
Redlich reminded 
us that a {\it canonical} rather than a grand-canonical 
ensemble with {\it exact conservation of quantum number locally} is required 
in the limit of strange particles $<$1/event, severely 
reducing the phase space available and thus explaining strangeness suppression in 
elementary reactions in a natural way (this is actually known since Hagedorn).
He also 
demonstrated the transition from one extreme to the other with a calculation of 
multistrange hyperon production as a function of the number of participants \cite{38,60}. 
This enhances the crucial importance of precise experimental information on the 
centrality dependence of hyperon production: NA57 \cite{57} has reported a first point with a 
decreased yield for $\overline{\Xi}^+$ hyperons in more peripheral collisions, but the error 
bar and the 
lack of more complete systematics prohibit firm conclusions at this stage. An 
experimental proof for a real onset behaviour in the production of strange hyperons (or 
other strange particles), either in impact parameter or mass number or beam energy 
dependence, would make a much more convincing case for strangeness production as a 
memory effect from quark matter formation than the popular model argument (if correct 
at all) that strangeness equilibration on the time scales available would only work with a 
preceding partonic scenario.

A final remark concerns $pA$ reactions. As was repeatedly reported during this conference 
\cite{61,62}, $pA$ collisions are also powerful to create additional strangeness relative to pp. 
As long as this is not very systematically investigated and clarified, including also the 
suspicious enhancement of strangeness up to the full level seen already at the AGS (see 
Fig.~\ref{fig:fig11}), the issue will remain very much controversial.

\subsection{Evidence for the chiral transition}

The medium properties around the phase boundary, where hadronization occurs, can 
experimentally be addressed by the unique tool of low mass dileptons. The instantaneous 
emission after creation and the absence of any final state interaction conserves the 
primary information within the limits imposed by the space-time folding over the 
emission period. In the low mass region, the thermal radiation is dominated by the 
decays of the light vector mesons $\rho$, $\omega$ and $\phi$. The $\rho$ is of particular interest, due to 
its direct link to chiral symmetry and its short lifetime of 1.3 fm/c; its 
in-medium behaviour around $T_c$ should therefore reflect chiral symmetry restoration, as 
proposed 20 years ago by Pisarski [63].

Experimentally, low mass dileptons are very much the domain of NA45/CERES, an 
electron pair spectrometer at the SPS which has probably taken more of my personal 
efforts over the years than any other experiment during my professional life. The results 
for PbAu at 160 AGeV, last updated at QM 1999, confirmed previous findings 
for S-Au (seen also by HELIOS 3 in the form of muon pairs). The combined 1995+1996 
data show an excess of electron pairs, in the mass region $>$ 0.2 GeV/c$^2$, of a 
factor of 2.9$\pm$0.3(stat.)$\pm$0.6(syst.) above the expectation from hadronic decay sources, setting in 
around 0.2 Gev/c$^2$; further findings are an unusually soft $p_T$-spectrum and a steeper 
than linear multiplicity dependence. More than 100 theoretical papers have appeared on 
the issue.
\begin{figure}[h]
\vspace{-.8cm}
\begin{minipage}[t]{80mm}
\includegraphics*[width=9cm]{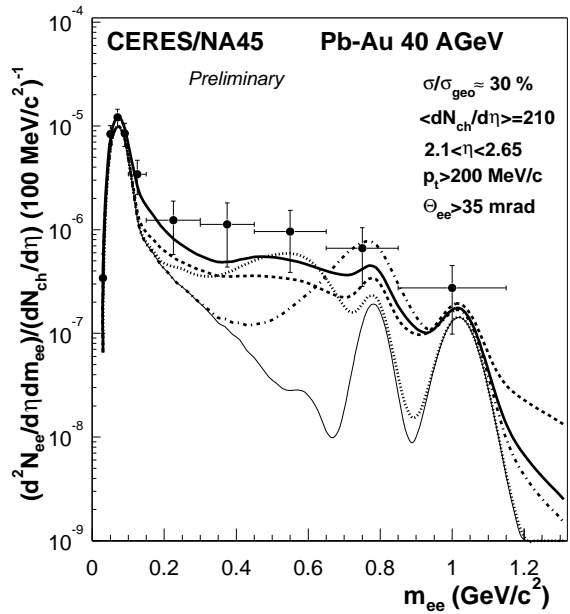}
\end{minipage}
\begin{flushright}
\vspace{-9.5cm}
\begin{minipage}[t]{65mm}
\caption{\footnotesize   Normalized invariant mass spectrum of $e^+e^-$ pairs at 40 AGeV in comparison to 
the sum of the known hadronic decay sources (thin solid line, without the $\rho$). The model 
calculations also shown [65] are based either on $\pi^+\pi^-$ annihilation with an unmodified $\rho$ 
(dashed-dotted), an in-medium dropping $\rho$-mass (dotted) and an in-medium spreaded 
$\rho$ (thick solid), or on $q\overline{q}$ annihilation (dashed) in the spirit of hadron-parton duality 
(see text). The figure is updated compared to the version shown at the conference [68,69].  }
\label{fig:fig12}
\end{minipage}
 \end{flushright}
\vspace{.5cm}
\end{figure}
 There seems to be a general consensus that one observes direct radiation, 
dominated by $\pi^+\pi^-$ annihilation, with a rate corresponding to an average 
temperature of T $\sim$ $T_c$ = 170 MeV \cite{23}. The temperature window contributing is about 120-220 MeV [23], 
implying that only $\sim$ 10 $\%$ of the observed yield is due to $q\overline{q}$ annihilation from the 
initial (deconfined) part \cite{23,64,65}. The shape of the mass spectrum seems to require a 
strong medium modification of the intermediate $\rho$. The main contenders for this are 
Brown-Rho scaling \cite{66}, shifting the mass, and a hadronic many-body calculation of the 
$\rho$ spectral density, spreading the width \cite{67}. The spread is so large that the whole 
spectrum can be described, as a parametrization, as if it were due to $q\overline{q}$ annihilation, in 
the spirit of hadron-parton duality \cite{67,23}. The relation to chiral symmetry restoration 
is there, but not straightforward \cite{66,67}; insight into the behaviour of the chiral partner 
$a_1$ would be highly desirable, as stressed in Gale's review \cite{20}.

CERES has now been upgraded with a TPC to obtain a better mass resolution, and this 
has also very much improved the hadron capabilities of the experiment. As the CERES 
report went along \cite{68} with one hadron result after the other, one of my wonderful 
former CERESian students asked, somewhat shocked, whether CERES made a phase 
transition from hadron-blind to electron-blind. The proof of the contrary with the 
preliminary electron pair data at 40 AGeV is shown in Fig.~\ref{fig:fig12} \cite{68,69}. A strong 
enhancement of 5.0$\pm$1.5(stat.) above the hadronic decay sources is again seen; within 
the limits of statistics and resolution (the experiment was not quite ready), there is 
consistency with the normalized data at 160 AGeV and with the model calculations for 
40 AGeV \cite{65}; the average temperature required is now reduced to 145 MeV \cite{23}. Better 
statistics than ever before, by a factor of 3-5, was obtained in the 2000 run at 160 AGeV, 
unfortunately discarding the multiplicity dependence. The mass resolution will be 
improved down to $\sim$ 2 $\%$. It remains to be seen whether in-medium effects can now be 
isolated for the $\omega$ and $\phi$ as well, in case of the latter also from comparing the decays 
into $e^+e^-$ and $K^+K^-$ within the same set-up. Due to the long lifetimes, the chances may be 
remote \cite{70} (ref. \cite{64} at this conference does not contain the dominating contribution 
from the decays in vacuum after freeze-out). Further running of CERES lies in the dark. 
Much improved data at a lower SPS energy and better insight into the multiplicity 
dependence at any energy seem almost mandatory. Unfortunately, the new experiment 
NA60, powerful for the $\omega$ and $\phi$ at higher $p_T$, will not become competitive for m $<$ 0.7 
GeV/c$^2$, due to the low $p_T$ cut inevitably connected with muon pair measurements.

\subsection{Expansion and Freeze-Out}

The transverse momentum distributions of the produced hadrons, the 
azimuthal aniso\-tropy $v_2$, and two particle interferometry are the experimental tools to probe the 
properties of the fireball in its final stage of thermal kinetic freeze-out, when all strong 
interactions between the constituents stop. The essential parameters to be discussed in 
this last chapter are the freeze-out temperature, the asymptotic velocity of the radially 
expanding fireball and the freeze-out density.

{\bf Transverse Momentum Distributions}

\begin{figure}[h]
\vspace{-.9cm}
\centerline{\includegraphics*[width=14cm]{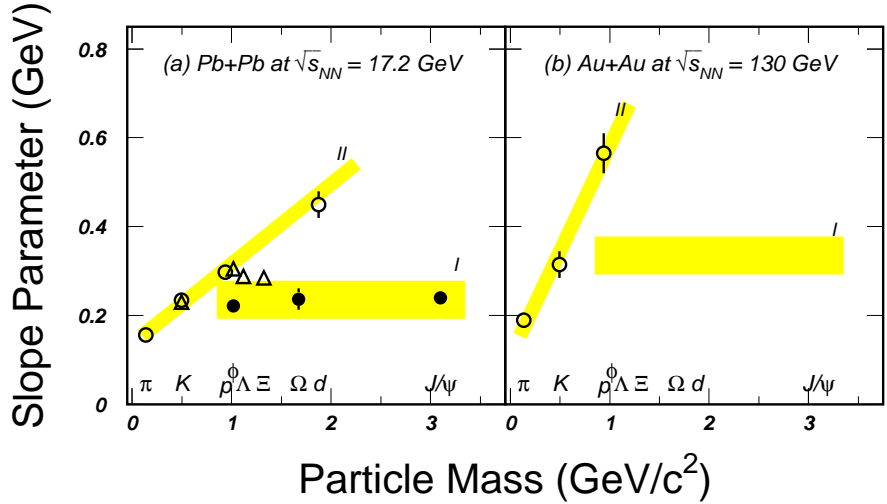}}
\vspace{-2.1cm}
\caption{\footnotesize Dependence of the transverse mass inverse slope 
parameter on the rest mass of 
the produced hadrons for the SPS ($\sqrt{s_{NN}}$ = 18 GeV, left) and for 
RHIC ($\sqrt{s_{NN}}$ = 130 GeV, 
right). The bands marked by I and II denote weakly resp. strongly interacting particles, 
see text (from [45]).}
\label{fig:fig13}
\vspace{-.5cm}
\end{figure}

\begin{figure}[h]
\vspace{-1.cm}
\centerline{\hspace{1cm}\includegraphics*[width=17cm]{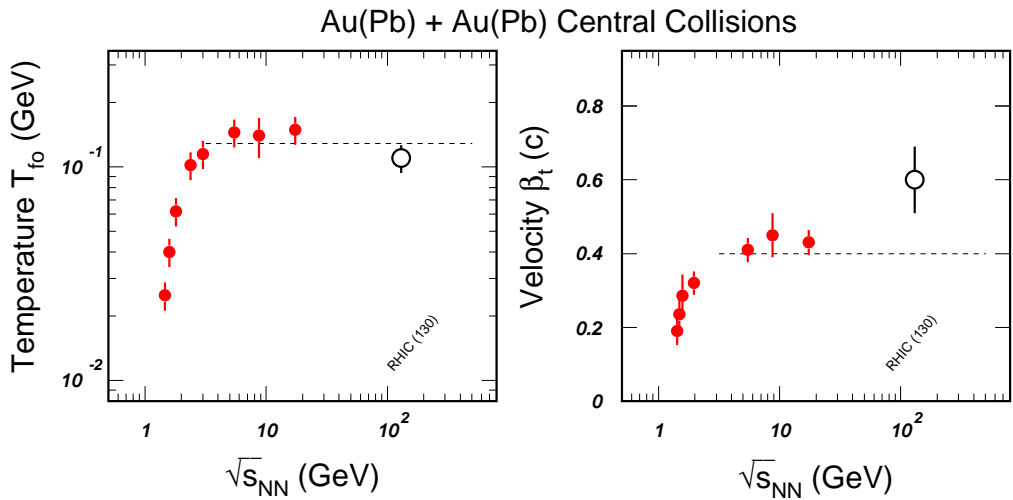}}
\vspace{-1.5cm}
\caption{\footnotesize Thermal-kinetic freeze-out temperature $T_{fo}$ and average 
collective transverse 
flow velocity $<$$\beta_t$$>$ as a function of center-of-mass energy $\sqrt{s_{NN}}$ (from [45]).}
\label{fig:fig14}
\vspace{-.5cm}
\end{figure}

New data on identified particle transverse momentum distributions have been reported 
at this conference for all energy regimes, including for the first time 40 AGeV at the SPS 
(from NA45 \cite{68} and NA49 \cite{71}) and, of course, RHIC (from STAR \cite{72}, PHENIX \cite{28} 
and BRAHMS \cite{73}); a summary of these data and their analysis in terms of the freeze-out 
parameters was presented by Nu Xu \cite{45}. The inverse slope parameters of the 
transverse mass distributions as a function of the rest mass of the produced hadrons are 
shown in Fig.~\ref{fig:fig13} with a direct comparison between full energy SPS and RHIC; the 
extracted values for the thermal freeze-out temperature $T_{fo}$ and the average collective 
transverse flow velocity $<$$\beta_t$$>$ as a function of the centre-of-mass energy are contained in 
Fig.~\ref{fig:fig14}, including AGS and 40 AGeV SPS data. A number of features are noteworthy. 
The increase of the inverse slope parameters with mass for the abundant particles $\pi$, $K$, 
$p$ etc. as visible in Fig.~\ref{fig:fig13} directly reflects a transversely 
expanding source in the 
spirit of roughly $T_{slope} = T_{fo} + const \cdot$$<$$\beta_t$$>$$\cdot m$ (const$\sim$0 
for $pp$ and $e^+e^-$). The collective expansion arises from 
the pressure gradient between the vacuum and the dense equilibrated matter which 
cools and dilutes until the interactions stop, reaching the asymptotic value of $<$$\beta_t$$>$. But 
what part of the system evolution contributes mostly to $<$$\beta_t$$>$? 
In Fig.~\ref{fig:fig13}, 
the slope parameters seem to fall into two categories: one (II), just discussed, showing 
the expansion, the other (I) with the $\phi$, $\Omega$ and $J/\psi$ (new at this 
conference) being flat. 
To the extent that the group (I) particles are all characterized by small interaction cross 
sections with the other hadrons of the system (``early freeze-out"), this has been taken as 
evidence 
that the plateau-like flow with a value of about 0.45 of the velocity of light in the 
region $\sqrt{s_{NN}}>$ 5 GeV (see Fig.~\ref{fig:fig14}) essentially develops in the late hadronic 
stage of the 
collision, while the contribution from the primordial part (at the top SPS energy) may be 
small (``soft" equation of state). 
The first RHIC point at $<$$\beta_t$$>$$ \sim$ 0.6, though still with 
some error (see the steeper slope in Fig.~\ref{fig:fig13} and Fig.~\ref{fig:fig14}), is therefore 
a further most 
remarkable and important new result from the initial round: is it evidence, like a 
memory effect, for a strong contribution from a preceding quark matter phase (``stiffer" 
equation of state at the higher initial temperature)? 
Future systematic data with 
reduced errors and values of the slope parameters for the $\phi$, $\Omega$ and $J/\psi$ will 
be crucial to confirm these first hints for a possible primordial flow at RHIC. Two particle 
interferometry will also be of use here (see below).

The freeze-out temperatures $T_{fo}$, plotted in the left part of Fig.~\ref{fig:fig14}, 
hardly need 
discussion. They also saturate at $\sqrt{s_{NN}}>$ 5 GeV, reaching a universal value of about 120 
MeV, and the first RHIC point does not seem to be very different from that, at least not 
within the present errors. 

\begin{figure}[h]
\vspace{-1.cm}
\begin{minipage}[t]{80mm}
\hspace{-1.3cm}\includegraphics*[width=9.5cm]{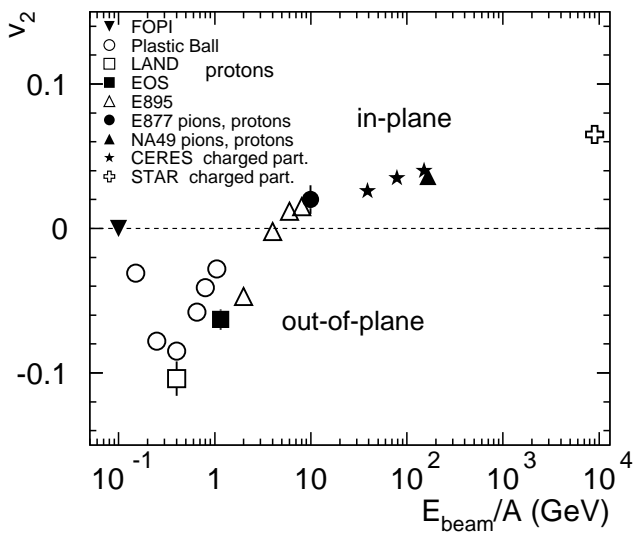}
\end{minipage}
 \begin{flushright}
\begin{minipage}[t]{60mm}
 \vspace{-8.5cm}
\caption{\footnotesize Azimuthal anisotropy $v_2$ relative to the reaction plane as a 
function of the beam 
kinetic energy for semi-central collisions of Pb or Au nuclei (from [68]).}
\label{fig:fig15}
\end{minipage}
\end{flushright}
\vspace{-3.7cm}
\end{figure}

Independent information on the degree of rescattering or thermalization with particular 
weight on the early time of the expansion is contained in the azimuthal anisotropy $v_2$. 
Still another most striking result from the first round at RHIC is the large value of 0.06, first 
found by STAR \cite{74} and 
then confirmed by the other experiments. A discussion on the systematics of $v_2$ is 
contained in Steinberg's review \cite{6} during this conference. The most comprehensive 
accumulation of data on $v_2$ was actually shown by Appelsh\"auser \cite{63}. Fig.~\ref{fig:fig15} 
reproduces 
these data, covering the whole beam energy regime from 0.1 AGeV at SIS to RHIC. 
Three new (still preliminary) points from NA45/CERES \cite{63} seem to suggest a rising 
trend with $\sqrt{s}$ such that the high point at RHIC is not necessarily jump-like. It remains 
to be seen whether $v_2$ is the ultimate quantity to convincingly confirm the existence of a 
primordial flow at RHIC.

{\bf Two Particle Interferometry}

\begin{figure}[h]
\vspace{-1.6cm}
\begin{minipage}[t]{80mm}
\includegraphics*[width=9.cm]{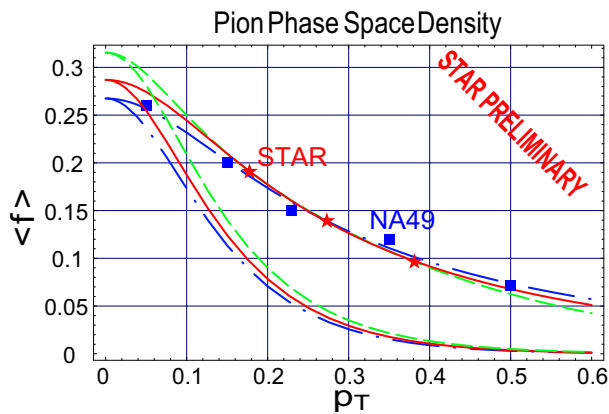}
\end{minipage}
\begin{flushright}
\begin{minipage}[t]{60mm}
\vspace{-6.7cm}
\caption{\footnotesize Pion phase-space density at freeze-out for $\sqrt{s_{NN}}$ = 18 GeV at the SPS (squares) and 
for $\sqrt{s_{NN}}$ = 130 GeV at RHIC (stars). A model description is also shown; the upper group of 
lines contains the influence of flow, the lower one does not. The associated freeze-out 
temperatures are rather low, 100/94/90 MeV for the 3 lines of each group (from [75]).}
\label{fig:fig16}
\end{minipage}
\end{flushright}
\vspace{-2.2cm}
\end{figure}

Small relative momentum correlations, known as HBT interferometry, have proven 
extremely useful to study the space-time structure of the fireball evolution in its last 
stage. New data were reported at this conference for 40 AGeV at the SPS (from 
NA45/CERES \cite{68} and NA49 \cite{71}) and, again of course, RHIC (from STAR \cite{75}); a review 
of the present situation was presented by Panitkin \cite{76}. The first results from RHIC can 
be summarized as follows. (i) Unusually large source sizes, proposed as a signature for 
quark matter formation \cite{77}, have so far not been observed. (ii) The ratio $R_\circ$/$R_s$ is, 
somewhat surprisingly, found to be $\leq$ 1 and decreasing with $k_T$. (iii) The spatially 
averaged 6-dimensional phase-space density $<$f$>$ of the pions, deduced from the HBT 
radii and the $\pi^-$ transverse mass distribution, confirms the hypothesis of a ``universal" 
phase space density at freeze-out \cite{78}. As shown in Fig.~\ref{fig:fig16}, the $p_T$-dependence of $<$f$>$ 
agrees for the data at RHIC and at the SPS (maybe too well), and the model description is clearly 
inconsistent with a static thermal source, but rather requires a Bose-Einstein 
distribution modified to include radial flow (with a fit value of $\beta$ = 0.58 at RHIC, 
consistent with the spectral analysis). The new results from NA45/CERES \cite{68} at 40 
AGeV and from E895 \cite{79} at their uppermost energy of 8 AGeV at the AGS also roughly 
agree with the data of Fig.~\ref{fig:fig16} at low $p_T$, but show increasing deviations towards smaller 
values of $<$f$>$ at higher $p_T$.

It appears then, as a joint conclusion of sections 2.3 and 2.5, that the fireball evolution 
from hadronization onwards until final freeze-out is essentially the same at the SPS and at RHIC, 
with the possible exception of an increased expansion rate at RHIC. 

\section{ Concluding Remarks}

I have written the summary of this conference in the style of an autark mini-review to 
enhance its usefulness. The physics conclusions are synonymous to the introduction of 
section 2 and therefore do not need to be repeated. However, I do 
have some afterthoughts. Twenty years ago we wanted to detect quark matter. We 
conceivably have seen first glimpses of it at the SPS and already now at RHIC. But we 
wanted more, namely quantitative physics: the equation of state as a function of 
temperature, the details of the phase- and the chiral transitions etc. etc. RHIC offers a 
huge spectrum of exciting new opportunities, and they will doubtlessly be used. But are 
we altogether on the right track? Have we properly explored the historical opportunities? 
I have some sympathy with B. M\"uller's theoretical summary at QM 1999 and 
C. Lourenco's experimental SPS summary at QM 2001. The SPS has 
presumably been the ideal machine for the phase {\it transition} region. Major points have, 
however, been left open which I touched upon in the respective sections and which 
concern all four surviving experiments NA45/CERES, NA49, NA57 and NA50 $\rightarrow$ NA60. 
One can only hope for the wisdom of all of us carrying responsibility, that we will find a 
proper balance between the needs at the SPS, at RHIC and at LHC to really get the 
optimum. 

{\bf Acknowledgements.}
I am grateful to many colleagues for critical discussions and useful support, in particular to A. Drees, 
K. Redlich, H. Satz, E. Shuryak and T. Ullrich. I am also indebted to A. Mar\'{\i}n for preparing 
the print version of the manuscript and to J.P. Wessels for his help with 
Figs.~\ref{fig:fig1} and \ref{fig:fig2}. I finally thank the organizers of QM2001 for their infinite 
patience with the manuscript.

\end{document}